\documentclass[aps, reprint, superscriptaddress, nofootinbib, showpacs, floatfix]{revtex4-1}

\usepackage{graphicx}  
\usepackage{dcolumn}   
\usepackage{bm}        
\usepackage{amssymb, amsmath}
\usepackage{slashed}   
\usepackage{color}
\usepackage[bookmarks]{hyperref}

\usepackage{tensor} 

\usepackage[usenames,dvipsnames,svgnames,table]{xcolor}

 \newcommand{\half}{\frac{1}{2}}

\def\beq{\begin{equation}}
\def\eeq{\end{equation}}
\def\M{\mathcal{M}}  
\def\mp{m_{\mathrm{Pl}}} 

\begin{document}

\widetext
\begin{flushright}
{\small Saclay-t19/013}\\
\end{flushright}

\title{Amplitudes' Positivity,  Weak Gravity Conjecture, and Modified Gravity}
\author{Brando Bellazzini}
\affiliation{Institut de Physique Th\'eorique, Universit\'e Paris Saclay, CEA, CNRS, F-91191 Gif-sur-Yvette, France}
\author{Matthew Lewandowski}
\affiliation{Institut de Physique Th\'eorique, Universit\'e Paris Saclay, CEA, CNRS, F-91191 Gif-sur-Yvette, France}
\author{Javi Serra}
\affiliation{Physik-Department, Technische Universit\"at M\"unchen, 85748 Garching, Germany}


\begin{abstract}
\noindent 

We derive new positivity bounds for scattering amplitudes in theories with a massless graviton in the spectrum in four spacetime dimensions, of relevance for the weak gravity conjecture and modified gravity theories. The bounds imply that extremal black holes are self-repulsive, $M/|Q|<1$ in suitable units, and that they are unstable to decay to smaller extremal black holes, providing an S-matrix proof of the weak gravity conjecture. We also present other applications of our bounds to the effective field theory of axions, $P(X)$ theories, weakly broken galileons, and curved spacetimes.

\end{abstract}


\maketitle

\section{Introduction}

The general properties of the S-matrix, unitarity, analyticity and crossing symmetry, imply dispersion relations for forward elastic scattering amplitudes, which in turn yield positivity bounds for amplitudes evaluated in the infrared (IR). They provide therefore non-trivial constraints on the coefficients of operators in the effective field theories (EFTs) that are used to calculate the amplitudes at low energy~\cite{Adams:2006sv,Bellazzini:2016xrt}. An EFT with operators entering the action with the  ``wrong'' sign cannot arise as the low-energy limit of a consistent ultraviolet (UV) theory satisfying the S-matrix axioms, and thus lives in the ``swampland.'' 
The proof of the a-theorem \cite{Komargodski:2011vj,Luty:2012ww} is perhaps the prime example of an application of these positivity bounds. 

In this paper, new amplitudes' positivities are derived for theories with a massless graviton in the spectrum, despite the fact that the forward elastic 2-to-2 scattering is singular (Coulomb singularity). These new positivity bounds, and the way we circumvent the graviton forward singularity, are extremely important because they allow us to address the swampland program of quantum gravity and modified gravity theories,  providing general and robust results. 

As a notable application, we study the Einstein-Maxwell theory,  i.e.~the low-energy EFT of an abelian $U(1)$ gauge theory coupled to gravity, and show that our positivity bounds imply certain inequalities among the leading higher-dimensional operators affecting the black hole's extremality condition (the minimal mass for which a charged black hole can exist, as opposed to a naked singularity). We show that extremal black holes of mass $M$ and  $U(1)$-charge $Q$ must satisfy $\sqrt{2}\mp |Q|/M > 1$, so that they are self-repulsive and are no longer kinematically forbidden from decaying into smaller extremal black holes. Moreover, these positivity bounds provide a proof of the mild form of the weak gravity conjecture (WGC) \cite{ArkaniHamed:2006dz}: extremal black holes are themselves charged states in the theory for which gravity is the weakest force. 

Another interesting application is for shift-symmetric scalars such as in the EFT of axions, $P(X)$ theories, and weakly broken galileons. The latter are found to have a tiny cutoff if they are to originate from a canonical microscopic S-matrix, $\Lambda_{UV} < \mathrm{few}\times (H^3\mp)^{1/4} \sim1/(10^7\mathrm{km})$, i.e.~orders of magnitude smaller than the strong coupling scale $\Lambda_3=(H^2\mp)^{1/3} \sim 1/(10^3 \text{ km})$.

Finally, we also discuss how our methods allow us to derive positivity bounds in a mildly curved de Sitter (dS) spacetime.

\section{Regulating the Forward Limit}
\label{sectionII}
 
The forward elastic amplitude of massless particles of polarizations labelled by $z_i$ is dominated by the universal Coulomb singularity 
\begin{equation}
\label{eq:forwarduniversal}
\M^{z_1 z_2}(s,t\rightarrow 0)= -\frac{s^2}{\mp^2 t}+O(s) \, ,
\end{equation}
because of the equivalence principle or, equivalently, because of factorization of the amplitude at the pole into the soft emission of an on-shell massless graviton, which has universal strength given by the reduced Planck mass $\mp$. Since the coefficient of $s^2$ in (\ref{eq:forwarduniversal}) would enter the dispersion relation in the forward limit for particles of any spin $z_i$ \cite{Bellazzini:2016xrt}, see (\ref{eq:dispersiverel}), the na\"ive application of the Cauchy integral theorem to $ \M^{z_1 z_2}(s,t\rightarrow 0)$  yields $\infty=\infty$, which is consistent, but admittedly not very informative. It is clear, however, that this divergence is due to long-distance physics, i.e.~vanishing exchanged momentum,  corresponding to the graviton probing arbitrarily large macroscopic distances even for large center-of-mass energy squared $s$. 
In any EFT the presumption is that the IR physics is known, therefore one should be able to track and resolve the source of the forward singularity. Indeed, we show below how to massage the dispersion relation into an effective, regulated expression $\infty-\infty=\mathrm{finite}>0$, returning something meaningful, free of ambiguity and in fact of a definite sign, which can be used for charting the swampland in gravitational theories. 

The key observation is that the Coulomb singularity is due to the infinite flat-space volume.  One would be tempted to regulate it by putting the system in a box (or perhaps in anti-de Sitter), however that would break Lorentz invariance and spoil the usual arguments that lead to positivity bounds. A compromise is enough for our purposes: we regulate the system by putting it on a cylinder.  That is, we compactify one spatial direction on a circle of length $L$, while the other three spacetime dimensions remain flat and infinite. In this way we can use 3D Lorentz invariance of the non-compact dimensions and scatter 3D asymptotic states, while at the same time getting  rid of the Coulomb singularity. Indeed,  there is no propagating massless graviton in $D=3$, hence no $s^2/t$-term for any finite value of $L$.%
\footnote{The tree-level exchange of a 3D auxiliary 2-tensor $g_{\mu\nu}$ does give rise to an $s^2/t$-term. However, since the associated singularity does not correspond to any physical particle, i.e.~there is no physical graviton in $D = 3$, higher-order corrections actually remove it. In App.~\ref{app1} we show that the 3D Einstein-Hilbert amplitude is indeed regular in the forward limit (in fact constant), therefore irrelevant and can be subtracted in the dispersion relation (\ref{eq:dispersiverel}). This should be contrasted to the $D=4$ case, where no correction can erase the $t$-channel pole.}
The 4D  graviton has not fully disappeared though, it has rather left three propagating avatars (on top of  a non-propagating auxiliary field $g_{\mu\nu}$ which gives rise to contact terms):
\begin{equation}
{\hat g}_{MN}\rightarrow \{ \sigma\,, \,\, V_\mu\,,\,\,\mbox{KK-modes} \} \, ,
\end{equation}
a massless dilaton $\sigma$, a massless (abelian) graviphoton $V_\mu$, and an infinite tower of Kaluza-Klein (KK) modes with masses $m_n^2\sim n^2/L^2$. In the limit $L\rightarrow \infty$, that we take at the end after isolating the diverging terms, one recovers the 4D dynamics we are interested in.%
\footnote{We stress that we are not discussing 3D toy models as done in a similar context in \cite{Cheung:2014ega,Chen:2019qvr}. Those references work in a truly $D=3$ setup, since in the IR spectrum there are neither the massless dilaton and the massless graviphoton, nor the massless 3D scalar $\Phi$ inside the 4D photon ${\hat A_M}$ (see Sec.~\ref{sec:EM}), nor the tower of massive (but light) KK modes, all of which are needed to correctly reproduce the 4D dynamics and thus affect the positivity bounds. Incidentally, the conclusion about the connection between neutrinos and electrons pointed out in \cite{Chen:2019qvr} seems premature, since the 3D IR spectrum of the compactification of the Standard Model is very different, in particular neutral light states other than neutrinos are abundant, and have non-minimal couplings.}

There is another advantage of compactifying to $D=3$, namely that asymptotic states all behave as massless scalars at high energies because the massless 3D little-group is trivial. This explains why the massless graviphoton is dual to a scalar field, and also explains why a massive KK-graviton decomposes into a massless scalar, a massless vector (dual to a massless scalar again) and a non-dynamical 2-tensor at high energy.

In the following (see Sec.~\ref{sec:EM}) we will be interested in scattering non-gravitational massless states, for example the 3D photon $A_\mu$ and the 3D scalar $\Phi$ that live inside the 4D photon ${\hat A}_M$, in the 4D Einstein-Maxwell theory reduced to 3D. Since there is no 3D massless graviton and the states are either gaped, non-propagating, or equivalent to simple scalars, we require a 3D Froissart-like bound
\begin{equation}
\label{eq:froissart3D}
\lim_{s\rightarrow \infty}|\M^{z_1 z_2}(s,t=0)/s^2|\rightarrow 0 \, , \qquad z_i=\Phi\,,\,\, A\,,
\end{equation}
where with a slight abuse of notation we are using $z_i$ to label now the scattered 3D states.
This is just the same assumption of polynomial boundedness that one accepts in 4D to derive dispersion relations and positivity bounds for the coefficient of e.g.~$(\partial\pi)^4$ or $(F_{\mu\nu} F^{\mu\nu})^2$, when 4D gravity is neglected or non-dynamical \cite{Adams:2006sv}.%
\footnote{In App.~\ref{app2} we discuss how our conclusions adapt to relaxing the Froissart bound (\ref{eq:froissart3D}). We show in particular that an asymptotic form of the WGC, i.e.~for very large extremal black holes, can still be proven even assuming no Froissart-like bound. In this context, we also discuss amplitudes' positivity for gravitational states, such as the 3D dilaton.}

For a gapped system, the Froissart bound becomes an actual theorem \cite{Froissart:1961ux,Martin:1965jj}, providing the asymptotic bound $|\M(s\rightarrow \infty)| < \mathrm{const}\,\cdot s\log^{D-2} s$ for $D\geq 3$ \cite{Chaichian:1987zt,Chaichian:1992hq}. One could thus even argue that (\ref{eq:froissart3D}) is automatically satisfied by giving a mass to the dilaton and to the graviphoton, then deriving the bound, and finally taking the massless limit, which is smooth for either field due to the abelian nature of the graviphoton. We will not commit to this view and content ourselves with assuming (\ref{eq:froissart3D}) and extracting the corresponding implications.

Therefore, under exactly the usual assumptions that lead to the familiar positivity bounds for systems of spin-0 and spin-1 massless particles in 4D, and repeating similar steps to those outlined in for example~\cite{Bellazzini:2016xrt}, we obtain a (provisional) dispersion relation for our IR-regulated 4D gravitational theory  
\begin{equation}
\label{eq:dispersiverel}
a^{z_1 z_2}= \frac{2}{\pi}\int_0^\infty \frac{ds}{s^3}\mathrm Im \M^{z_1 z_2}(s,t=0) >0 \,, 
\end{equation}
where the low-energy scattering amplitude for the 3D states $z_i$ is now regular in the forward elastic limit 
\begin{equation}
\label{eq:residue}
\M^{z_1 z_2}(s,t\rightarrow 0)=a^{z_1 z_2} s^2+\ldots \, .
\end{equation}

The dimensional reduction to 3D has left a universal contribution from gravitational zero and KK modes.
Each KK mode gives   
\begin{equation}
\label{KKestimate}
a^{z_1 z_2}_{KK}\propto  \frac{1}{L^2 \mp^4 m_{KK}} \propto \frac{1}{L \mp^4 |n|} \, ,
\end{equation}
where we used that the $n$th KK-mode mass is $m_{KK}\propto |n|\pi/L $. While each such contribution is subleading with respect to the terms we want to bound in the following sections, their sum is actually logarithmically divergent. In addition, zero-mode loops generate $s^{3/2}$-terms in the amplitude, which dominate over the $s^2$-terms at low energy, seemingly swamping again the information about $a^{z_1 z_2}$. In fact, these problems can be easily solved because the right-hand side of the dispersion relation (\ref{eq:dispersiverel}) reproduces  the same growth, so that these otherwise large terms cancel out between the two sides of (\ref{eq:dispersiverel}). Indeed, since the integrand itself in (\ref{eq:dispersiverel}) is positive by the optical theorem, schematically $\mathrm{Im}\M^{z_1 z_2}(s,t=0)=\sum_x |\M^{z_1 z_2\rightarrow x}|^2 \times \mbox{(phase space)}$, we can move to the left-hand side any contribution from intermediate states $x$ in $|\M^{z_1 z_2\rightarrow x}|^2$ and still get a positivity bound due to the remaining set of intermediate states. Specifically, we can move to the left-hand side the contributions from the intermediate IR states, such as the KK modes or anything that is calculable within the EFT (e.g.~IR loops, that is, the light multi-particle intermediate states). The zero- and KK-mode contributions get subtracted and one is left to calculate just the contact terms suppressed by the cutoff $\Lambda_{UV}$, that is, those that are generated by integrating out genuine UV states.
   
Just to illustrate this general point with a simple tree-level example, let us consider  $\Phi\Phi\rightarrow \Phi\Phi$ scattering with the exchange of a scalar state $S$ coupled to $(\partial\Phi)^2$,
\begin{equation}
\M^{\Phi\Phi}_{S}(s,t)= -\frac{2 c}{\mp^2 L}\left(\frac{s^2}{s-m^2_{S} +i\epsilon}+\mathrm{crossing}\right)\, ,
\end{equation}
where $c$ is a fixed $O(1)$ number. This contributes to $a^{z_1 z_2}$ in (\ref{eq:residue}) by an amount $a^{\Phi\Phi}_{S}=4 c/(\mp^2 L m^2_{S})$. The imaginary part (associated to the production of $S$) is
\begin{equation}
\mathrm{Im}\M^{\Phi\Phi}_{S}(s,t=0)=\frac{2\pi c}{\mp^2 L}m_{S}^4\delta(s-m_{S}^2)+\ldots \, ,
\end{equation}
precisely such that 
\begin{equation}
a^{\Phi\Phi}_{S} - \frac{2}{\pi} \int_0^\infty \frac{ds}{s^3} \mathrm Im \M^{\Phi\Phi}_{S}(s,t=0) = 0\,,
\end{equation}
as expected on general grounds. 

The KK-mode contributions to $a^{z_1 z_2}$ in (\ref{KKestimate}) actually arise at one loop, but the reasoning based on the optical theorem is completely general and works as in the previous example. This can be understood by discretizing the KK branch cut in a series of poles. Likewise for the contribution of the zero modes.%
\footnote{The tree-level exchange of an exactly massless dilaton is inconsequential because $s+t+u=\mathrm{const}$. If a stabilisation potential for the dilaton $\sigma$ is included, for example a mass term $\sim \sigma^2/L^2$, the tree-level subtraction also works for the dilaton pole.}.
The concrete details of how these contributions are subtracted are given in App.~\ref{app3}. Here we only note that the KK modes, which grow the ``extra'' dimension as seen from a low-energy 3D observer, reproduce nicely the 4D universal gravitational contribution to the RG running of $a^{z_1 z_2}$. This contribution is positive and as noted above would dominate the left-hand side of (\ref{eq:dispersiverel}). Since we can subtract it, which amounts to setting the renormalization scale at which $a^{z_1 z_2}$ is evaluated at the cutoff where UV and IR amplitudes are matched, our final dispersion relation properly captures the UV physics we are interested in.

All in all, our provisional dispersion relation (\ref{eq:dispersiverel}) is rearranged into a much more informative expression
\begin{equation}
\label{eq:dispersivesubtracted}
a^{z_1 z_2}-a^{z_1 z_2}_{KK, IR}=  \frac{2 }{\pi}\!\! \int_0^\infty \frac{ds}{s^3}\mathrm Im {\widetilde \M}^{z_1 z_2}(s,t=0)>0 \, ,
\end{equation}
where $\widetilde{\M}$ is the amplitude with the aforementioned gravitational zero- and KK-mode loop contributions subtracted. The left-hand side is therefore obtained by taking into account only the $s^2$-contributions to the elastic $z_1 z_2$-scattering due to the tree-level interactions with massless particles such as the graviphoton and the dilaton, as well as the UV generated contact terms, especially those from the auxiliary field $g_{\mu\nu}$.\footnote{Incidentally, the resulting contact terms can not be subtracted except for obtaining the useless relation $0=0$, since they do not correspond to any IR intermediate state that alone would satisfy (\ref{eq:froissart3D}).} 
The two sides (factor $L^{-1}$) of the subtracted dispersion relation (\ref{eq:dispersivesubtracted}), are not only finite for $L\rightarrow \infty$, but they are also positive because of the optical theorem.
We note that removing the IR modes from the positivity bound is always possible but is useful in practice only for UV completions that are not strongly coupled at $\Lambda_{UV}$, because it would become murky to assign what is IR (KK) and what is UV physics at around the scale $\Lambda_{UV}$. The subtracted dispersion relation is instead sharp and  useful for weakly coupled UV completions.

One general lesson is that gravity still has a finite effect on the positivity bounds even after removing the Coulomb singularity, due to the dilaton, the graviphoton,  and the auxiliary 2-tensor. This will be reflected in the new bounds derived for the explicit examples that we discuss next. 

\section{Einstein-Maxwell EFT}
\label{sec:EM}

\begin{figure}
\centering
\includegraphics[width=6cm]{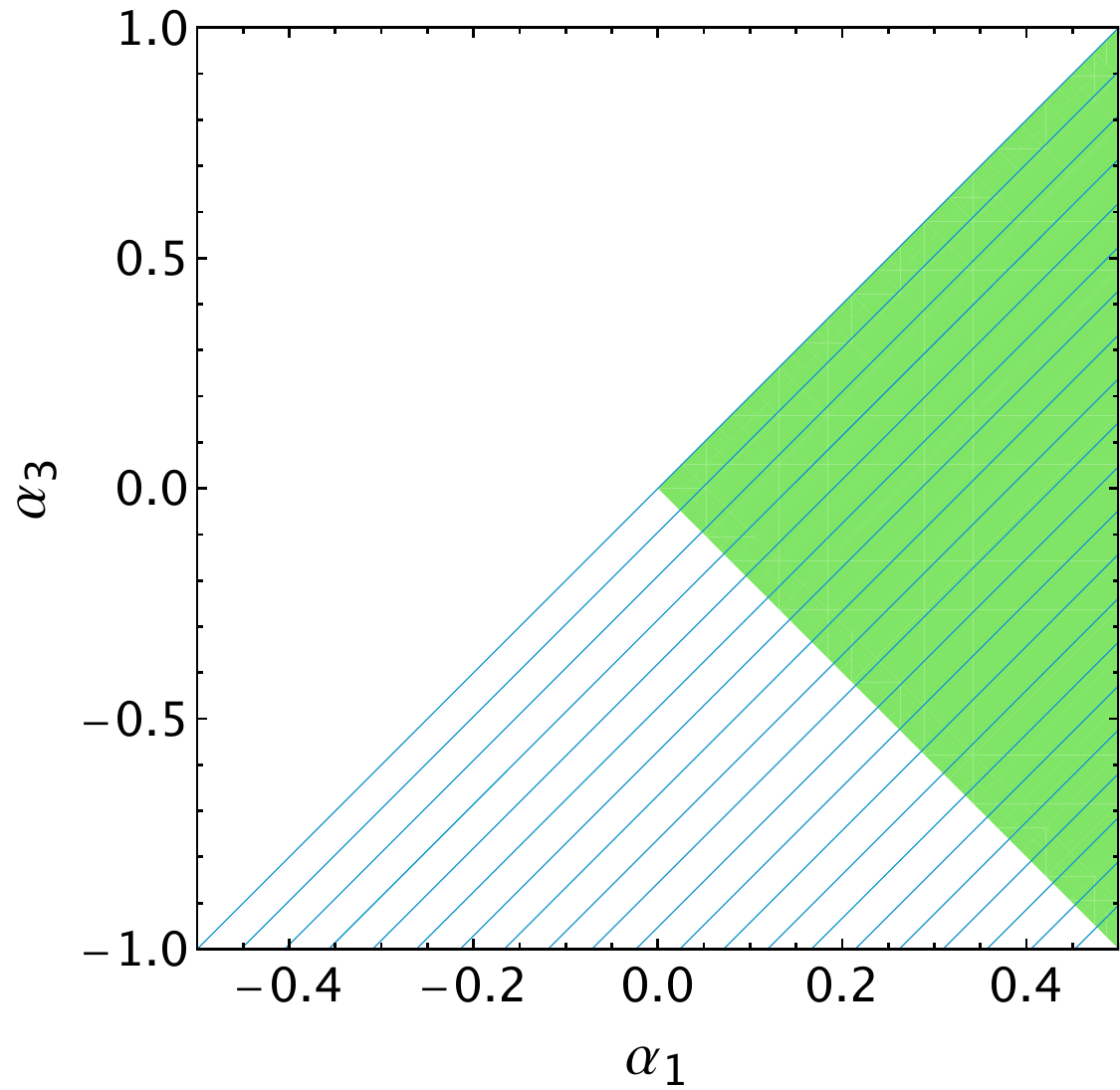}
\caption{Positivity bounds (\ref{Eq:positivityEM1}, \ref{Eq:positivityEM2}) require $\alpha_1$ and $\alpha_3$ to live inside the the smaller green wedge. The blue striped region is where extremal black holes are self-repulsive, $ |Q|>M/(\sqrt{2} \mp)$.}
\label{fig:1}
\end{figure}

In this section we focus on the important example of the Einstein-Maxwell EFT, whose leading 4D operators  are
\begin{align}
\label{eq:EMEFT}
S =& \int d^4 x \sqrt{|\hat g |} \left[ \frac{\mp^2}{2} \hat R - \frac{1}{4} \hat F^{MN} \hat F_{MN}    \right. \\ 
\nonumber
 &\left. + \frac{\alpha_1}{4 \mp^4}  \left( \hat F^{MN} \hat F_{MN}  \right)^2  + \frac{\alpha_2}{4 \mp^4}  \left( \hat{  \widetilde{  F}}{}^{MN}  \hat F_{MN}  \right)^2  \right. \\
 \nonumber
 &\left. +\frac{\alpha_3}{2 \mp^2}\hat F_{AB} \hat F_{CD} \hat W^{ABCD} \right] \, ,
\end{align}
where ${\hat W}^{ABCD}$  is the Weyl tensor and $ \hat{\widetilde{F}}{}_{MN}=\epsilon_{MNAB}{\hat F}^{AB}/2$. The dependence on the UV scale $\Lambda_{UV}$ that generates the $\alpha_i$ is absorbed into their definitions. These are the most general (parity preserving) four-derivative operators, up to field redefinitions \cite{Cheung:2014ega,Cheung:2018cwt}. 
In order to regulate the 4D forward limit and apply the positivity bounds (\ref{eq:dispersivesubtracted}), we compactify the $z$ direction as described in the previous section 
\begin{align}
d \hat s^2_{4} [ \hat g_{MN} ] &= e^{ \sigma} d s^2_{3} [g_{\mu\nu}]+ e^{- \sigma} \left( d z  + V_\mu dx^\mu \right)^2 \,,\\ 
{\hat A}_M  dx^M & = A_\mu dx^\mu + \Phi \, d z \, ,
\end{align}
where all of the 3D fields are functions only of $(t,x,y)$.
Focusing on  terms which contribute to the $s^2$ part of the amplitude for $\Phi\Phi\rightarrow \Phi\Phi$, $AA\rightarrow AA$, and $\Phi A\rightarrow \Phi A$  only,  the terms in the action that we must retain are
\begin{align}
\nonumber
&S = L \int d^3 x \sqrt{-g} \Bigg\{   \frac{\mp^2}{2} \left( R - \half ( \partial \sigma)^2 - \frac{1}{4} V^{\mu \nu} V_{\mu \nu } \right) \\ \nonumber
&   - \frac{1}{4} (1-\sigma)F^{\mu \nu} F_{\mu \nu }  - (1+\sigma)\half (\partial \Phi)^2 -\frac{1}{2}F_{\mu\nu}V^{\mu\nu}\Phi  \\
\nonumber
& + \frac{\alpha_1}{4 \mp^4}  \left( F^{\mu \nu} F_{\mu \nu}  + 2 (\partial \Phi)^2    \right)^2 + \frac{\alpha_2}{\mp^4}\left( \epsilon^{\mu \nu \rho} F_{\mu \nu} \partial_\rho \Phi   \right)^2\\
\nonumber
&  +\frac{\alpha_3}{\mp^2}\left[F_{\rho\mu} F_{\rho\nu} - \partial_\mu\Phi \partial_\nu \Phi \right] \left(R^{\mu\nu}-\frac{1}{3}g^{\mu\nu}R \right. \\
\nonumber
& \hspace{1.7in} \left. +\frac{1}{3}g^{\mu\nu}\square \sigma  -\nabla^\mu \nabla^\nu\sigma \right) \\
& -\frac{\alpha_3 }{\mp^2}F_{\mu\nu}\partial_\rho \Phi\left(\nabla^\rho V^{\mu\nu} + g^{\mu\rho} \nabla_{\alpha}V^{\nu\alpha} \right) \Bigg\}\,, 
\end{align}
where we made a field redefinition $A_\mu\rightarrow A_\mu+\Phi V_\mu$ to make gauge invariance manifest. The $g_{\mu\nu}$ propagates no degrees of freedom in $D=3$ and we can integrate it out, which is effectively equivalent to plugging the lowest-order equations of motion $R^{\mu\nu}-\frac{1}{2}g^{\mu\nu}R=T^{\mu\nu}/(L \mp^2)$ into the interaction terms, generating new contact terms. After removing $\square\sigma$ from the interactions with another field redefinition we get
\begin{align}
\nonumber
S &= L \int d^3 x \sqrt{-g} \Bigg\{   \frac{\mp^2}{2} \left(R - \half ( \partial \sigma )^2 - \frac{1}{4}  V^2 \right)  \\
& - \frac{1}{4}(1-\sigma)F^2  - \half (1+\sigma)(\partial \Phi)^2  -\frac{1}{2}F_{\mu\nu}V^{\mu\nu}\Phi \\
\nonumber
& + \frac{\alpha_1}{4 \mp^4}  \left( F^2  + 2 (\partial \Phi)^2    \right)^2 + \frac{\alpha_2}{\mp^4}\left( \epsilon^{\mu \nu \rho} F_{\mu \nu} \partial_\rho \Phi   \right)^2\\
\nonumber
&  +\frac{\alpha_3}{\mp^4}\left[F_{\rho\mu} F^{\rho\nu} F^{\mu\sigma}F_{\nu\sigma}-\frac{1}{2}F^4- (\partial\Phi)^4 +\frac{1}{2}F^2(\partial\Phi)^2 \right]\\
\nonumber
&  - \frac{\alpha_3}{\mp^2}\left(F_{\rho\mu} F^\rho{}_{\nu}-\partial_\mu\Phi \partial_\nu\Phi\right)\nabla^\mu\nabla^\nu\sigma \\
\nonumber
& -\frac{\alpha_3 }{\mp^2}F_{\mu\nu}\partial_\rho \Phi\left(\nabla^\rho V^{\mu\nu} + g^{\mu\rho} \nabla_{\alpha}V^{\nu\alpha} \right)   \Bigg\}\,,
\end{align} 
where $F^2=F_{\mu\nu}F^{\mu\nu}$, and the same for $V$. The associated subtracted forward elastic scattering amplitudes are
\begin{align}
\label{PhiPhiampfor}
{\widetilde \M}(\Phi\Phi\rightarrow\Phi\Phi)(s,t=0)= &  \frac{2s^2}{\mp^4 L}\left(2\alpha_1 -\alpha_3 \right) >0 \,,\\
{\widetilde \M}(AA \rightarrow AA)(s,t=0)= &  \frac{2s^2}{\mp^4 L} \left(2\alpha_1 +\alpha_3 \right) >0 \,,\\
{\widetilde \M}(\Phi A \rightarrow \Phi A)(s,t=0)= &  \frac{4s^2}{\mp^4 L} \alpha_2 >0 \,.
\end{align}
Therefore, the associated positivity bounds read
\begin{align}
\label{Eq:positivityEM1}
2\alpha_1 -\alpha_3 >0 \,,\\ 
\label{Eq:positivityEM2}
2\alpha_1 +\alpha_3 >0 \,,\\
\label{Eq:positivityEM3}
 \alpha_2 >0\,, 
\end{align}
or, equivalently, $\alpha_1 >|\alpha_3|/2$, $\alpha_2>0$. These new positivity bounds are one of the main results of this paper. In Fig.~\ref{fig:1} we show the region constrained in the $(\alpha_1,\alpha_3)$-plane, which provides non-trivial constraints on the 4D coefficients of the  EFT (\ref{eq:EMEFT}) that includes a massless graviton in the spectrum.

Remarkably, these bounds are stronger, meaning more general, than just pure 4D Euler-Heisenberg EFT without gravity \cite{Adams:2006sv,Bellazzini:2016xrt,Hamada:2018dde}, and carry extra information about $\alpha_3$, which enters the black hole extremality condition as we discuss in Sec.~\ref{sec:WGC}.  

Moreover, our homogeneous bounds (\ref{Eq:positivityEM1} -- \ref{Eq:positivityEM3}) are distinct from the order-of-magnitude causality bounds on $O(|\alpha_3|)$ \cite{Camanho:2014apa,Hamada:2018dde}, which are derived assuming positivity of time delay and tree-level UV completion of the Einstein-Maxwell lagrangian. See also \cite{Goon:2016une,Camanho:2014apa} for a nice discussion of detectability of superluminal propagation within an EFT, and how the Euler-Heisenberg lagrangian limit of the real-world QED avoids superluminality \cite{Drummond:1979pp}. 
 
It is interesting to compare the bounds (\ref{Eq:positivityEM1} -- \ref{Eq:positivityEM3}) with the 4D calculation of the same processes retaining the Coulomb singularity in the $t\rightarrow 0$ limit
\begin{align}
\mathcal{M}_{4D}^{\downarrow\downarrow}= &  -\frac{s^2}{\mp^2 t}- \frac{s}{\mp^2}+ \frac{2s^2\left(2\alpha_1 -\alpha_3 \right)}{\mp^4} \,,\\
\mathcal{M}_{4D}^{\uparrow\uparrow}= &  -\frac{s^2}{\mp^2 t}- \frac{s}{\mp^2}+ \frac{2s^2\left(2\alpha_1 +\alpha_3 \right)}{\mp^4} \,,\\
\mathcal{M}_{4D}^{\uparrow\downarrow}&=  -\frac{s^2}{\mp^2 t}- \frac{s}{\mp^2}+ \frac{4s^2 \alpha_2}{\mp^4} \,,
\end{align}
where the up and down arrows represent the two choices of real linear polarizations.%
\footnote{We use real linear polarizations because they correspond to crossing symmetric amplitudes \cite{Bellazzini:2015cra,Bellazzini:2016xrt}, up to the terms due to the Coulomb singularity.}
The lesson is that our 4D-regulated calculation, which works with 3D Lorentz invariance of the cylinder,  teaches us which finite parts we are allowed to retain for the positivity bounds: throw away the $s^2/t$ singularity, the finite $O(s)$ term, but retain precisely the $O(s^2)$ term. 

This immediately prompts us to expect a continuous set of positivity bounds associated with
arbitrary linear polarizations $|c_{1,2}\rangle=(c_{\theta_{1,2}}|\uparrow_{1,2}\rangle+s_{\theta_{1,2}}|\downarrow_{1,2}\rangle)$, namely 
\begin{equation} 
\label{Eq:positivityEM4}
\alpha_3(c_{2\theta_1}+c_{2\theta_2})+4\alpha_1c^2_{\theta_1+\theta_2}+4\alpha_2s^2_{\theta_1+\theta_2}>0 \, ,
\end{equation}
where $c_{\theta}=\cos\theta$ and $s_{\theta}=\sin\theta$.
We will check the bounds (\ref{Eq:positivityEM4}) from arbitrary linear combinations with our controlled, IR-regulated method in future work.

\section{Weak Gravity Conjecture and Extremal Black Holes}
\label{sec:WGC}

The leading higher-dimensional corrections $\alpha_{i}$ in the 4D Einstein-Maxwell EFT (\ref{eq:EMEFT}) modify the black hole extremality condition to \cite{Kats:2006xp}
\begin{equation}
\label{eq:boundBH}
\left( \frac{\sqrt{2}|Q|}{M/\mp}\right)_{\mathrm{extr.}} =1+\frac{4}{5}\frac{(4\pi)^2\mp^2}{M^2}(2\alpha_1-\alpha_3)>1\,,
\end{equation}
where $M$ is the black hole mass and $Q$ its charge (including the gauge coupling), and we work around $M\simeq Q \mp\sqrt{2}$.  
Remarkably, on the right-hand side of this expression one finds the same combination $2\alpha_1 -\alpha_3 $ bounded to be positive by  (\ref{Eq:positivityEM1}). Therefore, positivity bounds imply  a greater charge-to-mass ratio for extremal black holes  than in pure general relativity coupled minimally to an abelian $U(1)$ gauge theory.  The lighter the extremal black hole, the larger the charge-to-mass ratio. Extremal black holes within the validity of the 4D EFT, i.e.~whose Schwarzschild radius $r_s = M/4\pi \mp^2$ is larger than $1/\Lambda_{UV}$, are therefore self-repulsive. 
  
The positivity bound (\ref{Eq:positivityEM1})  implies the mild form of the WGC \cite{ArkaniHamed:2006dz}, which states that a consistent theory of quantum gravity must contain massive charged states in the spectrum with $ |q|> m/(\sqrt{2}\mp)$: the extremal black holes of (\ref{eq:boundBH}) are such states. 
As a result, the paradox of stable extremal black holes has evaporated, since extremal black holes are no longer kinematically forbidden to decay into smaller black holes. Indeed, an extremal black hole of mass $M$ and charge $Q$ cannot decay into states that all have larger mass-to-charge ratio, since the spectrum of masses and charges $(m_i, q_i)$ is constrained by $M>\sum_i m_i$ and $Q=\sum_i q_i$, whereas 
$\sum_i m_i=\sum_i |q_i|  m_i/|q_i| >M$ which would be a contradiction. This argument is evaded precisely by decay products that contain one smaller extremal black hole, which has smaller mass-to-charge ratio (\ref{eq:boundBH}) because of the positivity bound (\ref{Eq:positivityEM1}).

Since the same combination of EFT coefficients, $2\alpha_1~-~\alpha_3$,  enters the Wald entropy shift  \cite{Cheung:2018cwt,Hamada:2018dde}, our positivity bound  (\ref{Eq:positivityEM1}) implies a larger black hole entropy as well.%
\footnote{Positivity bounds of even higher-derivative terms \cite{Bellazzini:2015cra}  imply positive shift in the Kerr black hole entropy, too \cite{Reall:2019sah}.}
Notice, however, that the reverse is not true.  Requiring that the shift of the Wald entropy (of electrically charged black holes) is positive as a starting point \cite{Cheung:2018cwt} does not produce the same bounds on $\alpha_{1}$ and $\alpha_3$, and says nothing about $\alpha_2$; while (\ref{Eq:positivityEM1}) is reproduced by demanding a positive entropy shift \cite{Cheung:2018cwt,Hamada:2018dde}, the conditions (\ref{Eq:positivityEM2}) $2\alpha_1+\alpha_3>0$, (\ref{Eq:positivityEM3}) $\alpha_2 >0$, and (\ref{Eq:positivityEM4}) are not.  

\section{Bounds on Scalars}

Our IR-regulated positivity bounds can now be used to constrain scalar theories, e.g.~the EFT of axions or modified gravity theories, that have a massless graviton in the spectrum.%
\footnote{Positivity bounds applied to the Lorentz invariant EFT of massive gravity had a dramatic impact \cite{Bellazzini:2017fep}.}
Let us consider for instance
\begin{equation}
\label{PofX}
\mathcal{L}=-\frac{1}{2}(\partial\phi)^2+\frac{a}{4f^4} (\partial\phi)^4+\ldots \, .
\end{equation}
Without gravity, the positivity bounds would imply $a>0$ \cite{Adams:2006sv}, and one could expect the same bound to hold as long as $f\ll \mp$. However, what about the case $f\gg \mp$? In fact, even for the case $f\ll \mp$, things are not completely obvious in modified gravity theories. 
  
For example, in the cosmological context a $P(X)$ theory is usually considered with a decay constant $f$ not too far from $\sqrt{\mp H} \equiv \Lambda_2 \ll \mp$, where $H$ is the Hubble constant.
However, the limit $t\rightarrow 0$ is an IR limit where the exchanged momentum goes to zero, so that intermediate massless particles such as the graviton affect macroscopic distances.%
\footnote{Incidentally, this is why we believe the assumptions in \cite{Baumann:2015nta} are not fully justified, since the graviton is probing curved spacetime regions even at large $s$.}
Even if one regulates the IR limit with the largest scale in the problem and takes $t \sim H^{2}$, the graviton Coulomb singularity (\ref{eq:forwarduniversal}) would give a contribution to the forward scattering which is at least as large as the one that we want to retain, i.e.~$O(s^2/{\mp^2 H^2})=O(s^2/f^4)$, thus spoiling the argument that leads to $a>0$. The same issue was originally pointed out in the context of 4D massive gravity \cite{Bellazzini:2017fep}, where the intermediate transverse graviton competes with the galileon modes for the contribution to the $a^{\pi\pi}\sim m_g^2/\Lambda_3^6\sim 1/(\mp^2 H^2)$ for a graviton mass $m_g\sim H$.

In the case of 4D massless gravity studied in this paper, after compactifying to $D = 3$ there is no Coulomb singularity, while even if the series of gravitational KK modes (\ref{KKestimate}) is formally logarithmically divergent, it can actually be removed in the subtracted dispersion relation (\ref{eq:dispersivesubtracted}). The left-hand side of (\ref{eq:dispersivesubtracted}) is thus finite and dominated by the contact terms, and since neither the dilaton nor the graviphoton contribute at leading order to the $s^2$-term in the forward $\phi\phi\rightarrow \phi\phi$ scattering, we conclude that in fact 
\begin{equation}
a>0
\end{equation}
holds true in any weakly coupled UV completion of an EFT of the type (\ref{PofX}), such as $P(X)$ theories coupled to gravity, and in fact even for axions with $f\gg \mp$ should their cutoff $\Lambda_{UV} = g_* f$ be smaller than the Planck mass.

More interesting conclusions apply to weakly broken galileons \cite{Nicolis:2008in,Pirtskhalava:2015nla}. Let us consider for example 
\begin{equation}
\mathcal{L}=-\frac{1}{2}(\partial\pi)^2- \frac{1}{2\Lambda_3^3}(\partial\pi)^2\square\pi + \frac{1}{4\Lambda_2^4} (\partial\pi)^4+\ldots \,,
\end{equation}
where one can imagine the natural situation where $\Lambda_2\gg \Lambda_3$ since $(\partial\pi)^4$ weakly breaks the Galilean symmetry whereas  $(\partial\pi)^2\square \pi$ is an invariant. It was shown indeed that the hierarchy 
\begin{equation}
\label{Lambda23}
\Lambda_2^4 \simeq H^2 \mp^2 \,, \qquad \Lambda_3^3 \simeq H^2 \mp \, ,
\end{equation}
is stable under the loop corrections due to gravity that break the galileon symmetry \cite{Pirtskhalava:2015nla}, and in fact even larger values of $\Lambda_2$ such as $(H\mp^2)^{1/3}$ are in principle consistent. 

However, as estimated in \cite{Nicolis:2009qm,Bellazzini:2016xrt} and calculated in detail in \cite{Bellazzini:2017fep}, the scales $\Lambda_2$ and $\Lambda_3$ cannot be arbitrarily separated while keeping the cutoff $\Lambda_{UV}$ fixed in a theory without gravity, because the integrand under the dispersion relations (\ref{eq:dispersiverel}) is strictly positive and it gives the following beyond positivity bound~\cite{Bellazzini:2017fep}: 
\begin{equation}
\label{appi4D}
a^{\pi\pi}=\frac{1}{\Lambda_2^4} > \frac{2}{\pi}\int^{\Lambda_{UV}^2} \frac{ds}{s^3}\mathrm{Im}\M^{\pi\pi}(s)\propto \frac{1}{16\pi^2}\frac{\Lambda_{UV}^8}{\Lambda_3^{12}}\,.
\end{equation} 
We can now see that a similar bound survives even when the graviton is dynamical if the UV completion is assumed to be weakly coupled (at least up to $\Lambda_2$). Following the arguments of the previous sections, we can subtract the gravitational KK modes after the 3D compactification, and then extract  the following bound 
\begin{equation}
\label{appi3D}
\frac{1}{\Lambda_2^4 L} > \frac{2}{\pi}\int^{\Lambda_{UV}^2} \frac{ds}{s^3}\mathrm{Im}\widetilde{\M}^{\pi\pi}(s)> \frac{c}{16\pi^2}\frac{\Lambda^{8}_{UV}}{L\Lambda_3^{12}}\,,
\end{equation}
where in the last inequality we used the optical theorem and retained the contribution to the inelastic cross-section into two galileon KK modes $\pi_k$, $\sum_{k, m_k < \Lambda_{UV}}\sigma(\pi\pi\rightarrow \pi_{k}\pi_{k})$. The constant $c=O(10^{-4})$ is an inessential numerical factor resulting from integrating over the phase space and then along the branch cut. Loop corrections to the $s^2$-coefficient on the left-hand side of the dispersion relation are either very small or have been subtracted. The bound (\ref{appi3D}) nicely reproduces the scaling from the calculation without gravity in (\ref{appi4D}). As a consequence, the hierarchy (\ref{Lambda23}) between $\Lambda_2$ and $\Lambda_3$, which is stable because of symmetries, in fact requires an extremely small cutoff
\begin{equation}
\Lambda_{UV}< \left(H^3 \mp \right)^{1/4} \left(\frac{16\pi^2}{c}\right)^{1/8}\sim \frac{1}{10^{7}\,\mathrm{km}}
\end{equation}
in order to be consistent with the beyond positivity bound (\ref{appi3D}) that applies in a gravity theory. Vainshtein screening \cite{Vainshtein:1972sx} is no longer a valid EFT argument at scales shorter than $\Lambda_{UV}^{-1}$ \cite{Nicolis:2004qq,Bellazzini:2016xrt} because operators with arbitrarily more derivatives per field insertion become large, non-suprisingly, since new degrees of freedom are excited at $\Lambda_{UV}$.  

It would be interesting to apply these new and powerful beyond positivity bounds to other general and still structurally robust EFTs of modified gravity~\cite{Santoni:2018rrx}.

\section{Curved spacetime}

In this section we argue how positivity bounds can be extended to the case of spacetimes which are barely $dS_4$ (or $AdS_4$), as appears to be the case in our universe with a 4D cosmological constant $\Lambda_{4}$ that is very close to the scale of the neutrino masses.  One trivial way would be to assume that we can vary $\Lambda_4$ down to zero while the EFT coefficients we are interested in depending very little on such a change. We entertain instead another possibility, where $\Lambda_4$ is  held fixed and less relevant or even marginal operators (e.g.~Yukawa couplings)  are varied. 

It was shown in \cite{ArkaniHamed:2007gg} that the Standard Model coupled to gravity has a landscape of 3D vacua which is accessible by varying the properties of neutrinos (or the value of $\Lambda_4$, or both) by just $O(1)$. One could for example vary the mass of the lightest neutrino, or the type (Dirac vs Majorana), or the mass splitting squared $\Delta m_{12}^2$, etc.. As these parameters are varied one gets $dS_3$ or $AdS_3$ solutions (times the compact dimension) which are energetically favored over $dS_4$. Importantly, a flat 3D Minkowski solution (times the compact dimension) can be energetically favorable as well. This can be reached for example by varying $\Delta m_{12}^2$ from $8\cdot 10^{-5} \text{ eV}^2$ to $1.5\cdot 10^{-5} \text{ eV}^2$ for Majorana neutrinos \cite{ArkaniHamed:2007gg}. Tuning the theory with such a value, we can run again the positivity arguments derived in the previous sections to constrain the coefficients $a$  of, say, $a \, e^4(F_{\mu\nu})^4/\Lambda_{UV}^4$, which is obtained by integrating out massive charged states at the UV scale $\Lambda_{UV}$. Since the neutrinos are  neutral and weakly coupled, one natural expectation is that changing e.g.~$\Delta m^2_{12}$ while holding everything else fixed, will not dramatically backreact on the value of $a$, that is $a(\Delta m_{12}^2)=a( \Delta m_{12}^2|_{SM})+O(\delta \Delta m_{12}^2/\Lambda_{UV})^2$. Alternatively, if the see-saw scale $\Lambda_\nu$ for generating Majorana neutrino masses, $m_\nu\sim v^2/\Lambda_{\nu}$, is much higher than the scale $\Lambda_{UV}$ we are interested in, then the contribution from particles at $\Lambda_{\nu}$ to the parameter $a$ is quickly overrun by the physics at $\Lambda_{UV}$. One would thus conclude that $a>0$ even in $dS_4$ with a finite $\Lambda_4$, within an accuracy $O(\delta \Delta m_{12}^2/\Lambda_{UV}^2, \Lambda_{UV}^4/\Lambda_\nu^4)$. 

The same logic can be applied to bound modified gravity theories, for example $P(X)$ theories, assuming that changing e.g.~$\Delta m_{12}^2$ (or the neutrino physics at $\Lambda_\nu$) does not change the cosmological constant $\Lambda_4$ by orders of magnitude. For example, one could imagine that $\Lambda_4$ is obtained by tuning various UV parameters against each other, possibly associated with scales  even higher than $\Lambda_\nu$, so that they  would impact the change in the coefficient of the IR irrelevant operators even less. 

 We worked with a specific example related to the Standard Model coupled to gravity,  but the idea is general and can be easily adapted to other cases, adding for example particles that are not coupled to the Standard Model and yet contribute to the vacuum selection through their Casimir energies.
 
\section{Conclusions and Discussion}
  
In this paper we derived new amplitudes' positivities in quantum gravity in four dimensions. We showed how to regulate and subtract the gravitational Coulomb singularity in the forward elastic limit by putting the theory on a cylinder, using its 3D residual Lorentz invariance, and then restoring to 4D spacetime. This method allowed us to extract positivity bounds on the $s^2$-coefficient of the EFT amplitudes removing the $t$-channel graviton singularity in a controlled way.  Remarkably, the resulting positivity bounds are generically different than those obtained in flat space without gravity. This is due to the contribution to the amplitudes from the dilaton and the graviphoton (on top of the contact terms from the non-dynamical metric), which remain dynamical even on the cylinder and leave their finite gravitational footprint in the 4D limit. 
  
As an important application we studied the Einstein-Maxwell EFT and showed that the positivity bounds imply stronger inequalities than for the Euler-Heisenberg lagrangian. In turn, the bounds imply that extremal black holes have a charge-to-mass ratio larger than one, which is approached from above as the mass is increased. This provides an S-matrix proof of the mild form of the WGC since it implies that extremal black holes are self-repulsive, $|Q|>M$ in suitable units, and unstable to decay to smaller black holes. The amplitudes' positivity imply as well that the Wald entropy shift due to the leading higher-dimensional operators is always positive. 
 
In the context of the ``swampland'' program, these are perhaps somewhat negative results, since they lower the expectations that the WGC is useful to chart the landscape of consistent theories of quantum gravity.  We employed only very general, basic S-matrix principles, and yet we have been able to show that extremal black holes are no longer kinematically stable thanks to the higher-dimensional operators generated by ``any'' weakly coupled UV completion with a consistent S-matrix. Of course, it may be that string theory is the only UV completion with such a canonical S-matrix.%
\footnote{We note that the logical possibility exists that violation of our bounds points instead towards a fundamental obstruction to flat 3D compactifications, no matter how large the radius of the compact dimension is taken. }
  
Given our bounds (\ref{Eq:positivityEM1} -- \ref{Eq:positivityEM3}) on the Einstein-Maxwell EFT coefficients $\alpha_i$, one could try to follow the strategy of \cite{Cheung:2014ega,Chen:2019qvr,Andriolo:2018lvp}, that is, to see whether the established bounds imply specific constraints on microscopic QFT models with $U(1)$ massive and charged particles that are integrated out to generate the $\alpha_i$. Such a general program faces, however, an obstruction because in 4D there are charge-independent and UV-sensitive contributions to the $\alpha_i$ from graviton loops (or dilaton, graviphoton, and KK-mode loops in the IR-regulated theory on the cylinder), which are not calculable in the QFT (i.e.~they require knowledge of the details of the UV completion of quantum gravity). One could perhaps make some progress in this direction with the extra assumption that such purely gravitational UV contributions are somehow small \cite{Cheung:2014ega}.
  
Another future direction is the exploration of the effective theory of $p$-forms coupled to gravity and whether the extremality conditions for black branes are related to the positivity bounds in quantum gravity once higher-dimensional operators are included in the EFT.  The case of a zero form $\phi$, an axion, would be extremely important phenomenologically as the analog WGC could reveal an obstruction in taking transplanckian decay constants.

We considered other important applications in cosmology within the context of modified gravity, such as the positivity bound on the leading shift-symmetric operator for scalars coupled to gravity, like e.g.~axions, $P(X)$ theories, and galileons. In particular, we found that perturbative UV completions of galileons can be consistent with the (beyond) positivity bounds in a theory with a massless graviton only if the cutoff of the theory is at least as small as $\mathrm{few}\times (H^3\mp)^{1/4}$. We have also discussed how positivity bounds can be extended to dS spacetime with a small cosmological constant by varying e.g.~the properties of neutrinos.  
  

\subsection*{Acknowledgements}
We would like to thank Luca Martucci, Riccardo Rattazzi, Francesco Riva, Marco Serone, Francesco Sgarlata, and Filippo Vernizzi for useful discussions. We also thank Miguel Montero for interesting comments and Garret Goon for discussions on the scattering of gravitational states. We are especially grateful to Toshifumi Noumi, Clifford Cheung and Grant Remmen for discussions on the soft limit. BB thanks Sergei Dubovsky for lively discussions about the existence of analytic scattering amplitudes in 3D conical spaces. JS is supported by the Collaborative Research Center SFB1258 and the DFG cluster of excellence EXC 153 ``Origin and Structure of the Universe.'' ML acknowledges financial support from the Enhanced Eurotalents fellowship, a Marie Sklodowska-Curie Actions Programme, and the European Research Council under ERC-STG-639729, \emph{preQFT: Strategic Predictions for Quantum Field Theories}.

\subsection*{Conventions}
We use the $(-,+,+,+)$ metric convention where the 4D (3D) tensors (do not) have a hat, latin (greek) upper (lower) case indices run over 4D (3D) values, and where $\hat{R}^A{}_{BCD}=\partial_C \Gamma^{A}_{DB}+\ldots$, $\hat{R}_{AB}=\hat{R}^C{}_{ACB}$, $\hat{W}^{ABCD}=\hat{R}^{ABCD}-\mbox{traces}$. Field strength tensors are defined as $F_{\mu \nu} = \partial_\mu A_\nu - \partial_\nu A_\mu$, and the Levi-Civita tensor is defined as $\epsilon_{\mu_1 \cdots \mu_d}  = \sqrt{|g_d|}  \, \varepsilon [\mu_1 \cdots \mu_d]$, where $\varepsilon [\mu_1 \cdots \mu_d ] = \pm 1 , 0$ is the standard Levi-Civita symbol. We use natural units $\hbar = c \equiv 1$.


\appendix
\section{Forward limit and graviton pole}
\label{app1}

In this appendix we discuss some special features of $D=3$ gravitational scattering in the forward limit.  

In 3D flat space, the propagator of the metric fluctuations $h_{\mu\nu}$, say in harmonic gauge, seems to give rise to the offending $s^2/t$-term in elastic amplitudes at tree level.
This happens despite the fact that there is no massless graviton in the spectrum, the reason being that the forward limit does not actually put the internal $h_{\mu\nu}$ leg on a physical on-shell one-particle state. Indeed, $t=0$ is obtained in the physical kinematics when the exchanged momentum in the $t$-channel vanishes, $q=k_1-k_3=(0,\vec{k}_1-\vec{k}_3)\rightarrow (0,\vec{0})$, which is just a point of the light-cone $q^2=0$.  Such a momentum $q$ is not carried by a massless one-particle state, since it has no energy, but it rather corresponds to the soft scattering of a state with the same quantum numbers of the vacuum. Therefore, naively there is a singularity in the scattering amplitude at $t = 0$ that does not correspond to a particle on-shell. The situation is different for more general complex kinematics where 
\begin{equation}
t=-q^2\rightarrow 0 \,, \quad q\neq 0 \,,
\end{equation} 
which would correspond to an on-shell particle.  In this case the amplitude factorizes into the product of two 3-point amplitudes where all legs are now on-shell. In $D=3$, this non-forward $t=0$ kinematics is possible only when all momenta are parallel, since $k_i^2=k_1\cdot k_3=k_2\cdot k_4=0$. In this case we have that as $t\rightarrow 0$ also $s\rightarrow 0$,  implying that $s^2/t\rightarrow 0$, confirming the absence of a physical, on-shell graviton in the spectrum. 

Precisely because there is no propagating graviton in $D = 3$,  as opposed to the 4D case, higher-order corrections can -- and in fact do -- shift the pole in 3D \cite{Ciafaloni:1992hu,tHooft:1988qqn,Deser:1993wt,Zeni:1993ew}. One way to reproduce this result is by resumming the exchange of an infinite number of $t$-channel diagrams using the eikonal amplitude \cite{Kabat:1992tb} specialized to 3D \cite{Ciafaloni:1992hu,Deser:1993wt}. Focusing first on the pure Einstein-Hilbert contribution, one obtains
\begin{align}
\label{eik1}
\!\!\delta(b,s) \! = & \frac{1}{4 s}\int_{-\infty}^{+\infty}\!\!\!\frac{dq}{2\pi} e^{-iq b} \frac{1}{\mp^2 L}\frac{s^2}{q^2}=-\frac{s|b|}{8 L\mp^2} \, , \\ 
\label{eik2}
\!\!\!\!\mathcal{M}(s,t)\! = & -i2s\int_{-\infty}^{+\infty}\!\!\!\!\!\!\!db e^{ib q} e^{2i\delta(b,s)} =\frac{-16L \mp^2 s^2}{16L^2\mp^4 t+s^2}\! \, ,
\end{align}
where $q^2=-t$, and we dropped an irrelevant $i4\pi s\delta(q)$. The first-order term in a $1/\mp^2$ expansion reproduces the tree-level amplitude $-s^2/(t \mp^2L)$, but higher-order terms are even more singular for $t \to 0$, confirming the need to include all orders to arrive at the non-perturbative result (\ref{eik2}). 
The resummed forward amplitude does not grow with $s^2$, it actually goes to a constant, $-16L\mp^2$. This is irrelevant for the dispersion relation (\ref{eq:dispersivesubtracted}), and it can thus be subtracted from $\widetilde{\M}$, without spoiling the Froissart bound (\ref{eq:froissart3D}), very much like the case of the massive KK modes.  This should be contrasted with the 4D case, where the graviton pole is physical and no correction can possibly erase it \cite{tHooft:1987vrq,Kabat:1992tb}. 
 
The same result (\ref{eik2}) can be obtained by solving the motion of one of the scattered particles in the shock-wave spacetime generated by the other particle \cite{Deser:1993wt,tHooft:1988qqn}. This actually provides a nice geometrical interpretation of the result, since the space is flat except at the conical singularity where the particle is located: the resummed amplitude (\ref{eik2}) is clearly dominated by the classical scattering on the cone, by an angle $\theta$ satisfying $\sin\theta/2=\sqrt{s}/(4L\mp^2)$ (for small angle). This actually explains the singularity at $s= \pm 4L\mp^2 \sqrt{-t}$ in usual terms: it is generated by an on-shell particle, propagating through spacetime on two classical trajectories, each reaching one of the two points on the edge of the conical space that are identified. Hence, this generates an Aharonov-Bohm like effect due to the interference of the two different paths \cite{Deser:1993wt}. This is in complete analogy to the case of light bending in the background of cosmic strings in 4D; even though there is no static force, the particle accumulates two different phases from its two different trajectories around the string.

Importantly, in the large compact-dimension limit $L\rightarrow \infty$, one recovers flat Minkowsky space exactly. Nevertheless, taking the long-distance limit $t \to 0$ faster than decompactifying $1/L^2 \to 0$, the amplitude is regular (of course for energies within the EFT, $s \ll \Lambda_{UV}^2 \ll \mp^2$). This was possible because the leading gravitational effects can be resummed exactly in the compactified theory, whereas scattering in a gravitational theory with $D\geq 4$ non-compact dimensions does not grant as much control over the departure from exact Lorentz invariance, always present when scattering particles in gravity. In this regard, we recall that since $\Lambda_{UV} \gg 1/L$, the positivity constraints we derive in the main text bound 4D Wilson coefficients generated by the UV physics integrated out at distances $1/\Lambda_{UV}$, regardless of the limit $t \ll 1/L^2$.

In practice, we learn that the flat-space propagator for the graviton is a bad starting point for the forward scattering amplitude, and one should rather use the propagator in the background generated by the other particle. Incidentally, this explains why one does not get any subleading singularities at $t=0$ from graviton loops, e.g.~$s^2/\sqrt{-t}$, since $t=0$ is a regular configuration for the actual propagator. 

Once the leading order 3D theory is nicely behaving in the forward limit, we can safely add the perturbative corrections due the higher-dimensional operators in (\ref{eq:EMEFT}), generated by short-distance physics. Since these produce just contact terms in the elastic amplitude (\ref{PhiPhiampfor}), of the form $c \alpha_i s^2$, the eikonal resummation has no effect to leading order in the $\alpha_i$, namely 
\begin{equation}
\!\!\Delta\mathcal{M}_{\alpha_i}(s,t)=-i2s\int_{-\infty}^{+\infty}\!\!\!\!\!\!\!db \, e^{ib q} e^{2i\delta(b,s)} 2i\Delta_{\alpha_i}=c \alpha_i s^2 \, ,
\end{equation}
where $\Delta_{\alpha_i}=c \alpha_i s \delta(b)/4$. 
Another method that arrives at the same conclusions is described in \cite{Zeni:1993ew}.

We end this appendix with a few comments on the 3D flat-space polarizations in the harmonic gauge, $\partial_\mu \bar{h}^{\mu\nu}=0$ where $\bar{h}_{\mu\nu}\equiv h_{\mu\nu}-1/2 \eta_{\mu\nu}h$ and $h=h_\mu^\mu$. Looking at the matrix elements $\langle q| \bar{h}_{\mu\nu}|0\rangle \equiv \bar{\epsilon}_{\mu\nu}(q)e^{iq x}$ for $q\neq 0$ we get 
\begin{equation} \label{polar}
\bar{\epsilon}_{\mu\nu}(q)=  a \, q_\mu q_\nu+ b(s_\mu q_\nu+ s_\mu q_\mu)+c s_\mu s_\nu \, ,
\end{equation}
where $q^2=0$, $q_\mu \epsilon^{\mu}_\nu=0$, and  $s_\mu$ is a spacelike unit vector orthogonal to $q$ and to $\bar{q}$, the latter being another null vector such that $\bar{q}\cdot q=-1$, e.g. 
\begin{equation}
\!\!q^\mu=(E,E,0)\,,\,\, \bar{q}^\mu=\frac{1}{2E}(1,-1,0)\,,\,\, s^\mu=(0,0,1)\,. 
\end{equation}
In this way the metric can be written as $\eta_{\mu\nu}=-(q_\mu \bar{q}_\nu+ q_\nu \bar{q}_\mu)+s_\mu s_\nu$, from which it follows that 
\begin{equation}
\bar{\epsilon}_{\mu\nu}(q)=  a q_\mu q_\nu+ b(s_\mu q_\nu+ s_\mu q_\mu)+c (\eta_{\mu\nu}+q_\mu \bar{q}_\nu+ q_\nu \bar{q}_\mu ) \,.
\end{equation}
When the graviton polarization $\epsilon_{\mu\nu}=\bar{\epsilon}_{\mu\nu}-\bar{\epsilon}\eta_{\mu\nu}$ is contracted with a conserved and symmetric energy-momentum tensor, $q_\mu T^{\mu\nu}=0$, one gets $\epsilon_{\mu\nu} T^{\mu\nu}=0$, confirming that on-shell polarizations, such as those in the numerator of the propagator at the pole, give rise to contact terms only. 

Besides, in the harmonic gauge there remains a residual gauge symmetry $h_{\mu\nu}\rightarrow h_{\mu\nu}+\partial_{\mu}\xi_{\nu}+\partial_{\nu}\xi_{\mu}$  with $\square\xi_\nu=0$, that one can actually use to set to zero all polarizations. This is accomplished by choosing 
\begin{equation}
\xi_\mu= \left(\alpha q_\mu +\beta \bar{q}_\mu +\gamma  s_\mu\right) e^{iqx} \, ,
\end{equation}
which gives
\begin{align}
\delta \bar{\epsilon}_{\mu\nu} ( q ) =i 2\alpha \, q_\mu q_\nu +i\beta s_\mu s_\nu  +i\gamma (s_\mu q_\nu+s_\nu q_\mu ) \,,
\end{align}
and with $\alpha=i a/2$, $\beta=i c$ and $\gamma=ib$, cancels the expression in (\ref{polar}). 
In this decomposition, however, it was important that $q$ was a non-trivial null vector, rather than a vector that eventually vanishes in the forward limit, $q\rightarrow 0$ (approached from a spacelike direction). In the latter case, one is dealing with a large gauge transformation. That is why, in the harmonic gauge, the tree-level forward amplitude is still singular in the forward limit. Importantly, this is an artifact of the tree-level approximation. In the presence of a (lightlike) source, one can define a pure-gauge, piece-wise (i.e.~singular) metric as in~\cite{Deser:1993wt}, for which the graviton propagator is actually different than in flat space. The resulting (non-perturbative) forward amplitude grows less than $s^2$ \cite{tHooft:1988qqn,Deser:1993wt} and can be subtracted, as we have seen above. 
    
Finally, another avenue to further confirm our findings would be to remove all together the metric fluctuations, up to a choice of topology, by compactifying one further dimension down to a 2D torus, leaving in the physical spectrum two massless dilatons, the scalars from the 4D photon, and the (bound states of) KK modes.  None of these states are expected to generate a $t$-channel singularity, but only contact terms in the forward limit.  We leave the explicit analysis of $D = 2$ for the future. 

\section{Relaxing the Froissart bound and running coefficients}
\label{app2}

In this appendix we discuss how the positivity constraints are affected when relaxing the assumption of the Froissart-like asymptotic bound (\ref{eq:froissart3D}). We proceed first by dropping altogether the assumption of polynomial boundedness, to next increase gradually the bound until reaching (\ref{eq:froissart3D}).

\begin{figure}
\centering
\includegraphics[width=8.5cm]{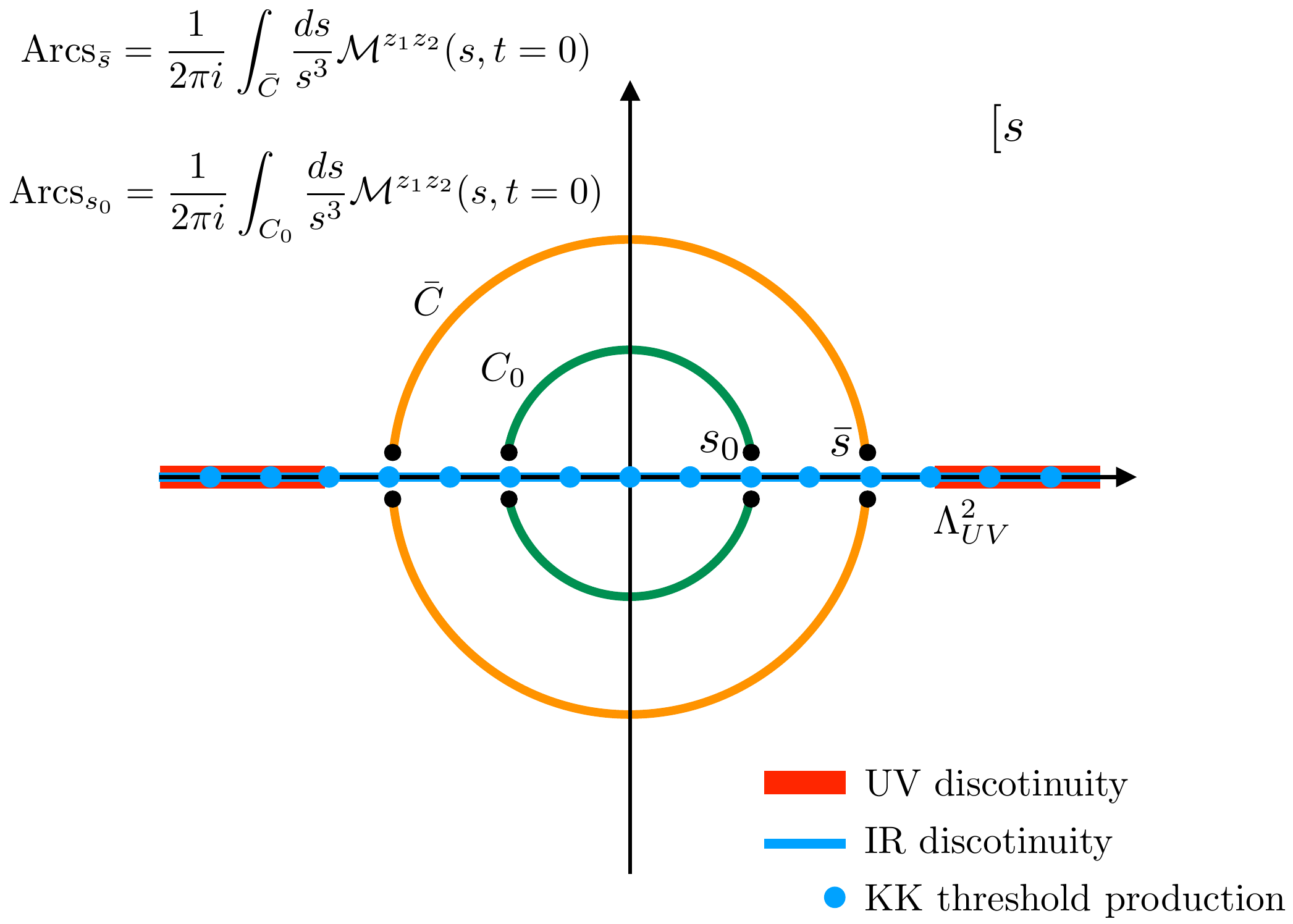}
\caption{Complex $s$-plane and singularity structure in the 4D theory compactified to 3D. The (double) arcs upon which the forward amplitude is integrated, see (\ref{arcs}), are also depicted.}
\label{fig:2}
\end{figure}

Assuming nothing about the asymptotic behavior of the forward elastic amplitude, we can still compare the integrals over (double) arcs at finite radius in the complex $s$-plane at $t=0$, as depicted in Fig.~\ref{fig:2}, namely 
\begin{equation}
\label{arcs}
\mathrm{Arc}_{s_0}-\mathrm{Arc}_{\bar s}=\frac{2}{\pi}\int_{s_0}^{\bar s}\frac{ds}{s^3}\mathrm{Im}\M^{z_1 z_2}(s,t=0) >0 
\end{equation}
for $\bar s > s_0$. This means that the Wilson coefficients run at $s_0$ are larger than those at $\bar s$, i.e.~the $\beta$-function for those coefficients selected by the forward amplitudes are negative, growing towards the IR. Here we define the Wilson coefficients by taking into account the leading logarithmic correction to the IR amplitude, 
\begin{align}
\M^{z_1 z_2}(s,t\!=\!0) &= a^{z_1 z_2}(\bar s)s^2 \\
& \,\,\, + \beta_a s^2 \tfrac{1}{2} \left(\log (s/\bar s) +\log (-s/\bar s) \right) \, .\nonumber
\end{align}
Equivalently, $\beta_a=d a^{z_1 z_2}(\bar s)/d\log \bar s$ from the RG equation $d\M^{z_1 z_2}/d\log\bar s = 0$. In the Einstein-Maxwell EFT of Sec.~\ref{sec:EM}, this corresponds to 
\begin{equation}
\beta_{2\alpha_1 \pm \alpha_3} <0 \,,\qquad \beta_{\alpha_2}< 0\,.
\label{betaEM}
\end{equation}
Therefore, positivity of the Wilson coefficients 
$(2\alpha_1\pm \alpha_3)$ and $\alpha_2$ evaluated at $s_0$ is guaranteed asymptotically, for $s_0/\Lambda_{UV}^2$ taken sufficiently small, corresponding to evaluating the extremality condition for very large black holes, $r_s \gg 1/\Lambda_{UV}$. In turn, this implies the WGC for such black holes as a consequence of unitarity and causality only, with no reference to a Froissart-like bound. Note that (\ref{betaEM}) was actually shown via explicit calculation in \cite{Deser:1974cz}.

We  move now to examine the next-to-weakest assumption on the asymptotic amplitude, namely 
\begin{equation}
\label{eq:weakerfroissart3D}
\lim_{s\rightarrow \infty}|\M^{z_1 z_2}(s,t=0)/s^2|\sim O\left( 1/(L \Lambda_{UV}^2 \mp^2) \right) \, ,
\end{equation}
which is the worst possible case compatible with no graviton in the spectrum and which reduces to the Froissart-like bound for $\mp\rightarrow \infty$. 
Despite the very weak assumption, we can still put positivity bounds directly on 
$a^{z_1 z_2}(\bar s \simeq \Lambda_{UV}^2)$ up to corrections of order $O(1/(L\Lambda_{UV}^2 \mp^2))$, which are 
good enough if the UV contributions to $a^{z_1 z_2}$ are expected to be larger, which it is often the case. For example, looking at the scattering of gravitational modes such as the dilaton, we find 
\begin{equation}
\widetilde{\M}^{\Phi \sigma}(s,t\!=\!0)=-\widetilde{\M}^{A \sigma}(s,t\!=\!0)=\alpha_3 \frac{s^2}{\mp^4L} \, ,
\end{equation}
and therefore $\alpha_3$ cannot be arbitrarily large, being bounded in fact by $|\alpha_3|<O(\mp^2/\Lambda_{UV}^2)$. This nicely reproduces the causality bound of \cite{Camanho:2014apa} but in a more controlled setting. 
Incidentally, this scaling reproduces the power counting of QED after integrating out a heavy electron: $\Lambda_{UV}=m_e$, $\alpha_3 \sim (g/4\pi)^2 (\mp/\Lambda_{UV})^2$ thus smaller than  $O(\mp^2/\Lambda_{UV}^2)$, and $\alpha_{1,2} \sim g^2 (g/4\pi)^2 (\mp/\Lambda_{UV})^4$, such that $\alpha_{1,2} \gg  |\alpha_3|$ as long as $\Lambda_{UV} \ll g \mp$, implying the WGC for any black hole within the EFT, not just asymptotically.
 
The next slightly stronger assumption is requiring (\ref{eq:froissart3D}) for scattering non-gravitational modes, yet (\ref{eq:weakerfroissart3D}) still holds for the gravitational ones. This again implies the WGC from the inequality $2\alpha_1-\alpha_3>0$ that can be derived as in Sec.~\ref{sec:EM}. In some cases, however, e.g.~the case of extended (and exact) supersymmetry in $D=4$ when the distinction between photons and gravitons is not possible, only the previous assumptions may be valid. This is no surprise, the positivity bounds are IR sensitive, and by changing the IR (additional massless fields are present in this cases, the Einstein-Maxwell EFT in isolation is no longer valid in any finite energy range), the bounds may change as well, very much like they have changed from pure Euler-Heisenberg theory to Einstein-Maxwell theory once gravity is turned on with a finite $\mp^2$.
Finally, we note that the asymptotic Froissart-like condition (\ref{eq:froissart3D}) is satisfied in string theory for tree-level scattering of open strings in $D\geq 4$, see e.g.~\cite{Adams:2006sv}. Including the Regge behaviour, associated with (loops of) closed strings in $D\geq 4$, might suggest instead the behavior (\ref{eq:weakerfroissart3D}) yet further suppressed by a factor $(g/g_{UV})^2$, $g_{UV}$ being associated to states that are ÒstronglyÓ coupled, in the sense that $g < g_{UV}$. It is however unjustified at this point to extrapolate these asymptotic limits at $t=0$ to $D=3$. 

\section{Gravitational zero and KK modes}
\label{app3}

In this appendix we discuss how the universal loop contributions from zero and KK modes can be subtracted from our dispersion relation. 

Let us then perform explicitly the relevant one-loop calculations. Consider the lagrangian 
\begin{equation}
\mathcal{L} = -\frac{1}{2}(\partial\Phi)^2 - \frac{c}{2L\mp^2} (\partial\Phi)^2 \sigma_{n}^2\, ,
\end{equation}
where $\sigma_{n}$ is one of the dilaton's $n$th KK mode (working with a real field), which contributes to $a^{\Phi\Phi}$ at one loop by an amount 
\begin{equation}
\label{KKexample}
a^{\Phi\Phi}_{\sigma_n} = \frac{c^2}{8\pi L^2 \mp^4 m_{\sigma_n}} \, .
\end{equation}
This is precisely matched by the integral of the cross-section 
\begin{equation}
\sigma(\Phi\Phi\rightarrow \sigma_n\sigma_n) = \frac{c^2\sqrt{s}}{16\mp^4L^2} \, ,
\end{equation}
over the KK branch cut, namely 
\begin{equation}
a^{\Phi\Phi}_{\sigma_n} - \frac{2}{\pi} \int^\infty_{4m_{\sigma_n}^2} \frac{ds}{s^2} \, \sigma(\Phi\Phi\rightarrow \sigma_n\sigma_n) = 0 \, .
\end{equation}
In this way, we can remove all of the KK loops. This procedure is also equivalent to working with a subtracted amplitude, e.g.
\begin{align}
{\widetilde \M}^{z_1 z_2}(s,t\!=\!0) &= \M^{z_1 z_2}(s,t\!=\!0) \\
&\quad - d \left( s^{3/2} \mathrm{arccot}\sqrt{4m_{\sigma_n}/s}+\mathrm{crossing}\right) \, , \nonumber
\end{align}
with $d=c^2/(8\mp^4L^2)$ in the specific one-loop example above. This corresponds to removing dilaton pair-production from the right-hand side of (\ref{eq:dispersiverel}), via de optical theorem (massless external states)
\begin{equation}
\label{opticaltheorem}
\mathrm{Im} \M^{z_1 z_2}(s,t=0) = s \sum_{x} \sigma_{z_1 z_2 \rightarrow x}(s) \, ,
\end{equation}
picking $x = \sigma_n \sigma_n$. Besides, notice that the asymptotic bound (\ref{eq:froissart3D}) is still respected by ${\widetilde \M}^{z_1 z_2}$.

Analogously, the zero-mode loops generate non-analytic terms of the type 
\begin{equation}
\frac{b}{L^2\mp^4}\left( s^{3/2}+(-s)^{3/2} \right) \,.
\end{equation}
However, in the limit $L\rightarrow \infty$ these decrease faster than the contributions that we want to bound, which scale instead as  $1/L$, see e.g.~(\ref{PhiPhiampfor}). In any case, this type of IR non-analytic terms can be subtracted as well \cite{Chen:2019qvr} by working again with a subtracted amplitude 
\begin{equation}
{\widetilde \M}^{z_1 z_2}= \M^{z_1 z_2} - \frac{b}{L^2\mp^4} \left( s^{3/2}+(-s)^{3/2} \right) \, ,
\end{equation}
which corresponds to removing the intermediate $x=\sigma_0\sigma_0$ from the sum over intermediate states under the dispersive integral (\ref{opticaltheorem}). The resulting low-energy amplitude is dominated by the $s^2$-terms that we want to bound. Finally, higher powers of $s$ do not affect the dispersion relation and therefore can be retained. 

More generally, one can subtract, selectively, any channel by using the optical theorem (\ref{opticaltheorem}), and up to any desired energy, as long as the Froissart bound (\ref{eq:froissart3D}) is satisifed. Since the imaginary part is bounded by $s^2$ so it is each individual positive contribution.  One can also be less selective and just subtract all IR contributions at once, by integrating in the complex $s$-plane along the arcs reported in Fig.~\ref{fig:2} in App.~\ref{app2} (in this case the dispersive integral starts from a finite $s$ value, $\bar{s} \neq0$, which can be taken somewhat smaller than the cutoff $\Lambda_{UV}^2$ of the EFT).

Finally, it is instructive to see how the KK-mode loops we have discussed reproduce the logarithmic running discussed in App.~\ref{app2}. As seen in (\ref{KKestimate}), each KK mode gives a finite IR contribution to $a^{z_1 z_2}$, $\Delta_n a^{z_1 z_2} = d/(8 \pi^2 L \mp^4 |n|)$. This follows e.g. from the explicit example in (\ref{KKexample}), after taking into account that $m_{\sigma_n} \propto |n| \pi/L$. Such KK contribution is associated with an IR branch cut starting at $s = 4 m_n^2$, see Fig.~\ref{fig:2}. Therefore, the difference between two arcs at $\bar s$ and $s_0$ is precisely given by the KK modes whose mass is within the range $s_0<4m_{n}^2< \bar s$. Therefore it follows that
\begin{align}
\mathrm{Arc}_{\bar{s}} &= a^{z_1 z_2}(\Lambda_{UV}^2) + \sum_{|n| > \frac{L \bar{s}}{4\pi}}^{|n|<\frac{\Lambda_{UV} L}{4\pi}} \Delta_n a^{z_1 z_1} \nonumber \\
& \simeq a^{z_1 z_2}(\Lambda_{UV}^2) + \frac{2d}{8 \pi^2 L \mp^4}\log(\Lambda_{UV}^2/\bar{s})\, ,
\end{align}
i.e.~subtracting the KK modes as discussed in Sec.~\ref{sectionII} is nothing but considering the running coefficients at (or rather near to) the matching scale $\Lambda_{UV}$ where the EFT is generated. Smaller arcs, that is including more KK modes in the calculation, corresponds to looking at larger distances where $a^{z_1 z_2}(s_0)$ is necessarily larger than $a^{z_1 z_2}(\Lambda_{UV}^2)$. Finally, it is even possible to work with a (double) arc of radius $\bar{s} > \Lambda_{UV}^2$, as long as we have a perturbative UV completion $\Lambda_{UV} \ll \mp$, e.g. where the zero and KK modes of photons and gravitons exist even above $\Lambda_{UV}^2$.

\bibliographystyle{utphys}
\bibliography{3dscattering}

\end{document}